\documentclass[11pt]{article}

\input bynd.rty

\begin{document}
\bibliographystyle{unsrt}

\centerline{\bf \huge From new states of matter to}
\vskip 0.2in
\centerline{\bf \huge a unification of light and electrons
}

\vskip 0.3in
\centerline{\bf Xiao-Gang Wen
\footnote{http://dao.mit.edu/~wen}
}
\vskip 0.2in
\centerline{Department of Physics, Massachusetts Institute of Technology}
\centerline{Cambridge, Massachusetts 02139}
\vskip 0.3in
\begin{abstract}
For a long time, people believe that all possible states of matter are
described by Landau symmetry-breaking theory.  Recently we find that
string-net condensation provide a mechanism to produce states of matter beyond
the symmetry-breaking description.  The collective excitations of the
string-net condensed states turn out to be our old friends, photons and
electrons (and other gauge bosons and fermions).
This suggests that our vacuum is a string-net condensed state.
Light and electrons in our vacuum have a unified origin --
string-net condensation.
\end{abstract}
%


\section{Introduction}

Quantum theory of condensed matter was dominated by two main themes. The first
one 
is Fermi liquid theory.\cite{L5620} The second theme is  Landau
symmetry-breaking theory.\cite{L3726,LanL58}  Fermi liquid theory is a
perturbation theory around a particular type of ground states -- the states
obtained by filling single-particle energy levels.  Fermi liquid theory
has very wide applications.  It describes metals, semiconductors, magnets,
superconductors, etc.  Landau symmetry-breaking theory provides a deep insight
into phase and phase transition. It points out that the reason that different
phases are different is because they have different symmetries.  A phase
transition is simply a transition that changes the symmetry.  Not so long ago
Landau symmetry-breaking theory is believed to describe all possible phases,
such as crystal phases, ferromagnetic and anti-ferromagnetic phases,
superfluid phases, etc., and all of the phase transitions between them.

Condensed matter theory is a very successful theory. It allows us to
understand properties of almost all forms of matter.  As a result, one starts
to get a feeling of completeness, and a feeling of seeing the beginning of the
end of the condensed matter theory. However, through the researches in last 20
years,
a different picture starts to
emerge. It appears that what we have seen is just the end of beginning.  There
is a whole new world ahead of us waiting to be explored.

A peek into the new world is offered by the discovery of fractional quantum
Hall (FQH) effect.\cite{TSG8259} Another peek is offered by the discovery of
high $T_c$ superconductors.\cite{BM8689} Both phenomena are completely beyond
the two themes mentioned above. Rapid and exciting developments in FQH effect
and in high $T_c$ superconductivity resulted in many new ideas and new
concepts.  Looking back at those new developments, it becomes more and more
clear that, in last 20 years, we were actually witnessing an emergence of a
new theme in condensed matter physics. The new theme is associated with new
states of matter and new class of materials.  This is an exciting time for
condensed matter physics. The new paradigm may even have an impact in our
understanding of fundamental questions of nature.

Why FQH effect starts a new theme condensed matter physics?  We know that
there are many different FQH states, but they all have the same symmetry. So
we cannot use Landau symmetry-breaking theory to describe different orders in
FQH states.  It was proposed that FQH states contain a new kind of order ---
topological order \cite{Wrig,Wtoprev}. Topological order is new since it has
nothing to do with symmetry breaking, long range correlation, or local order
parameters.  None of the usual tools that we used to describe a phase applies
to topological order.  Despite this, topological order is not an empty concept
since it can be described by a new set of tools, such as the number of
degenerate ground states \cite{HR8529,WNtop}, quasiparticle statistics
\cite{ASW8422,BW9033,R9002}, and edge states \cite{H8285,Wedgerev}.  

Amazingly, FQH effect is not the first evidence that indicates the presence of
the third new theme.  The first sign showed itself 150 years ago even before
the first two themes were introduced.  

According to Landau symmetry-breaking theory, the gapless collective
excitations in a state are determined by the symmetry-breaking order in the
state.
Those excitations correspond to various waves (the fluctuations of order
parameter) in the state.  If particles organize themselves into a lattice, the
resulting crystal order breaks translation symmetry and leads to sound waves in
crystals.  The ferromagnetic order  that breaks the spin rotation symmetry give
rise to spin waves.  In fact, all known waves are originated this way from
symmetry breaking, except for the electromagnetic wave described by the Maxwell
equation! 

\begin{figure}[tb]
\centerline{
\includegraphics[scale=0.8]{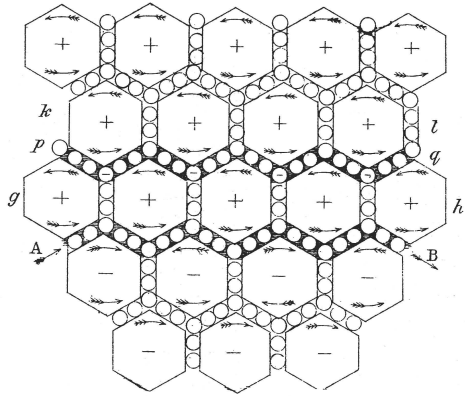}
}
\caption{
Maxwell's ether: a mechanical model that might give rise to Maxwell
equation and electromagnetic waves.
}
\label{MaxwellEther}
\end{figure}

When Maxwell equation was first introduced, people firmly believed that any
wave must be correspond to motion of something.  So people want to find out
what is the origin of the Maxwell equation?  The motion of what gives rise
electromagnetic wave?  Maxwell himself have tried to invent a mechanical model
so that the electromagnetic wave correspond to the motion of the mechanical
parts in the model (see Fig. \ref{MaxwellEther}).\cite{M6051} However,
Maxwell's mechanical model does not really work.  Furthermore, according to
Landau symmetry-breaking theory, fluctuating order parameter can never produce
waves that are described by Maxwell equation.  Thus if we believe that Landau
symmetry-breaking theory describe all possible states of matter (\ie all
possible organizations of particles), we will conclude that there is no way to
produce electromagnetic wave by properly organizing particles.  

However, after the discovery of the FQH states and the associated topological
order, we find that Landau symmetry-breaking theory does not describe all
possible organizations of particles.  So there is still hope.  The Maxwell
equation may arise from a new kind of  organizations of particles that are
beyond Landau symmetry-breaking theory. Those new organizations of particles
will correspond to new states of matter or a new class of materials.  This
motivates us to revisit the old question: what organization of particles can
give rise to Maxwell equation?  How to find those new states of matter?  How
to make those new materials?

In addition to the Maxwell equation, there is an even stranger equation, Dirac
equation, that describes wave of electrons (and other fermions).  Electrons
have Fermi statistics. They are fundamentally different from the quanta of
other familiar waves, such as photons and phonons, since those quanta all have
Bose statistics.  To describe the many-electron wave, the amplitude of the wave
must be anticommuting Grassman numbers, so that the wave quanta will have Fermi
statistics. Since electrons are so strange, few people regard electrons and the
electron waves as collective motions of something. People accept without
questioning that electrons are fundamental particles, one of the building
blocks of all that exist.  However, from a condensed matter physics point of
view, all low energy excitations are collective motion of something. If we try
to regard photons as collective modes, why cann't we regard electrons as
collective modes?  So we can ask a similar question for electrons: what
organization of particles can give rise to Dirac equation and Fermi statistics?
Are Fermi statistics and gauge structure fundamental laws of nature, or are
they emergent phenomena and a natural consequence of a particular organization
of simple bosonic particles?

\section{New states of matter from string-net condensations}

\begin{figure}[tb]
\centerline{
\includegraphics[scale=0.15]{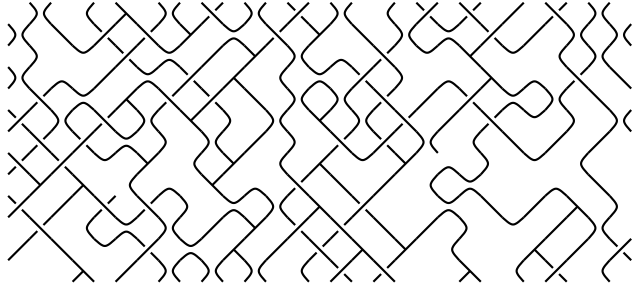}
}
\caption{
A quantum ether: The fluctuation of string-net give rise to
electromagnetic waves (or light). The ends of strings give rise to electrons.
}
\label{stringnetS}
\end{figure}

A recent study provides an answer to the above questions.
\cite{LWstrnet,LWuni,LWqed} We find that if particles form large strings and
if those strings form a liquid state, then the collective motion of the such
organized particles will correspond to waves described by Maxwell equation and
Dirac equation.  
The strings in the string liquid are free to join and cross each other. As a
result, the strings look more like a network (see Fig.  \ref{stringnetS}).
For this reason, the string liquid is actually a liquid of string-nets, which
is called string-net condensed state.\footnote{The particles that form the
string-nets are bosons. So in the string-net picture, the Maxwell equation
and Dirac equation
emerge from \emph{local} bosonic models. The electric field and the magnetic
field in the Maxwell equation are called gauge fields. The string-net liquids
demonstrate how gauge fields emerge from local bosonic models.  Many closely
related earlier works led to such a picture of the Maxwell equation.  String
structures appear in the Wilson-loop characterization\cite{W7445} of gauge
theory. The Hamiltonian and the duality description of lattice gauge theory
also reveal string structures.\cite{KS7595,BMK7793,K7959,S8053}.  Lattice gauge
theories are not local bosonic models and the strings are unbreakable in
lattice gauge theories.  String-net theory points out that even breakable
strings can give rise to gauge fields.\cite{HWcnt} So we do not really need
strings. Bosonic particles themselves are capable of generating gauge fields
and the associated Maxwell equation.  This phenomenon was discovered in several
bosonic models\cite{FNN8035,BA8880,Wlight,MS0204} before realizing their
connection to the string-net liquids.  The connection between ends of strings
and Fermi statistics is a more recent realization. \cite{LWsta}}

\begin{figure}[tb]
\centerline{
\includegraphics[width=2.5in]{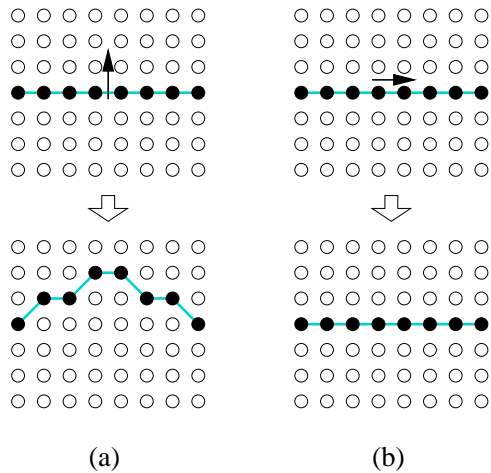}
}
\caption{
A hand-waving way to understand why fluctuations of condensed string-net have
only two transverse modes.
(a) A transverse motion of a string results in a new state and leads to
a collective excitation.
(b)  A motion along the string does not result in any new states. Such a motion
does not lead to any collective excitations.
}
\label{strvib}
\end{figure}

But why the waving of string-nets produces waves described by the Maxwell
equation? We know that the particles in a solid organized into a regular
lattice pattern.  The waving of such organized particles produces a compression
wave and two transverse waves.  The particles in a liquid have a more random
organization.  As a result, the waves in liquids lost two transverse modes and
contain only a single compression mode.  The particles in a string-net liquid
also have a random organization, but in a different way.  The particles first
form string-nets and string-nets then form a random liquid state. Due to this
different kind of randomness, the waves in string-net condensed state lost the
compression mode and contain two transverse modes.  Such a wave (having only
two transverse modes) is exactly the electromagnetic wave described by the
Maxwell equation.

To see how electrons appear from string-nets, we would like to point out that
if we only want photons and no other particles, the strings must be closed
strings with no ends.  The fluctuations of closed strings produce only
photons.  If strings have open ends, those open ends can move around and just
behave like independent particles.  Those particles are not photons. In fact,
the ends of strings are nothing but electrons.

How do we know that ends of strings behave like electrons?  First, since the
waving of string-nets is an electromagnetic wave, a deformation of string-nets
correspond to an electromagnetic field.  So we can study how an end of a string
interacts with a deformation of string-nets.  We find that such an interaction
is just like the interaction between a charged electron and an electromagnetic
field. Also electrons have a subtle but very important property -- Fermi
statistics, which is a property that exists only quantum theory.  Without their
Fermi statistics, the electrons in atoms would have a very different
organization. All atoms would have very similar chemical properties and behave
like a noble gas.  Amazingly, the ends of strings reproduce this subtle quantum
property of Fermi statistics.\cite{LWsta,LWstrnet}  Actually, string-net
liquids explain why Fermi statistics should exist.

We see that string-nets naturally explain both light and electrons.  In other
words, string-net theory provides a way to unify light and
electrons.\cite{LWuni,LWqed} So, the fact that our vacuum contains both light and
electrons may not be a mere accident. It may actually suggest that the vacuum
is indeed a string-net liquid.  

Here, we would like to point out that there are many different kinds of
string-net liquids.  The strings in those different liquids may have different
numbers of types and may join in different ways.  For some string-net liquids,
the waving of the strings does not correspond to light and the ends of strings
are not electrons.  Only one kind of string-net liquids give rise to light and
electrons.  On the other hand, the fact that there are many different kinds of
string-net liquids allows us to explain more than just light and electrons.  We
can find a particular type of string-net liquids which not only gives rise to
electrons and photons, but also gives rise to quarks and
gluons.\cite{Wqoem,LWstrnet} The waving of such type of string-nets corresponds
to photons (light) and gluons. The ends of different types of strings
correspond to electrons and quarks. It might even be possible to design a
string-net liquid that produces all elementary particles!  In this case, the
ether formed by such string-nets can provide an origin of all elementary
particles.\footnote{So far we can use string-net to produce almost all
elementary particles, expect for two: the $SU(2)$ gauge boson that is
responsible for the weak interaction and the graviton that is responsible for
the gravity.}

Because of so many different string-net condensed
states, the string-net condensed states are much richer
than symmetry-breaking states.  So the new theory of string-net condensation
may even be a richer theory
than the Landau symmetry-breaking theory.  The study of string-net condensed
states and the associated new states of matter may become a new theme in
theoretical condensed matter physics.

\section{A rotor model with string-net condensation}

\begin{figure}[tb]
\centerline{
\includegraphics[scale=0.4]{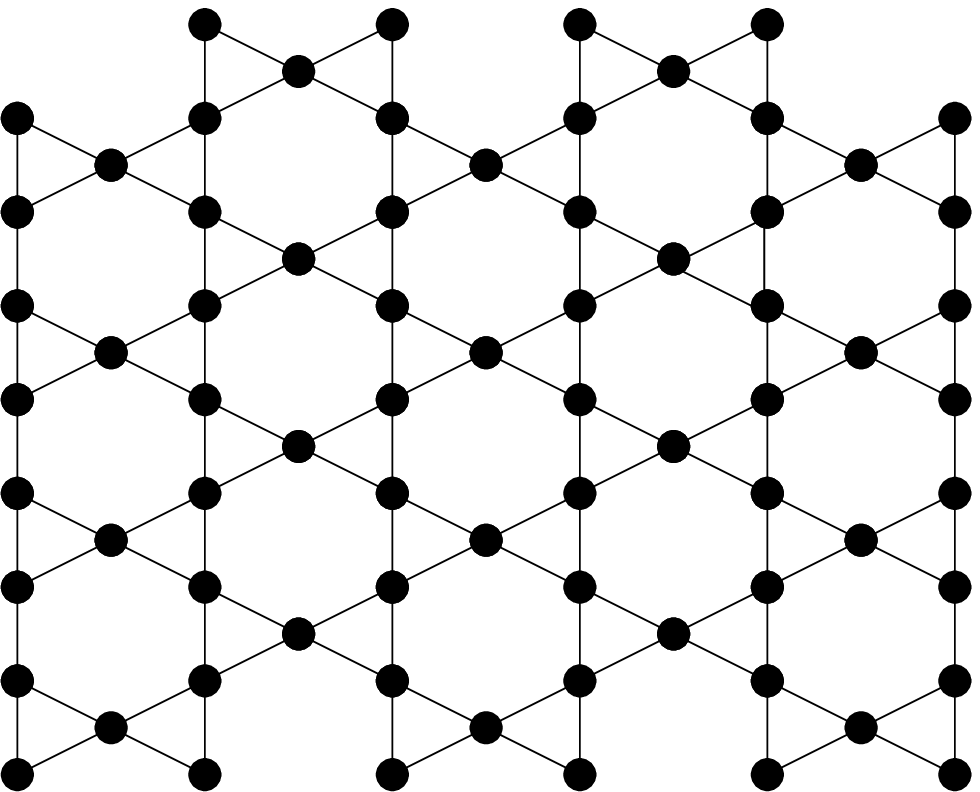}
\hfil
\includegraphics[scale=0.4]{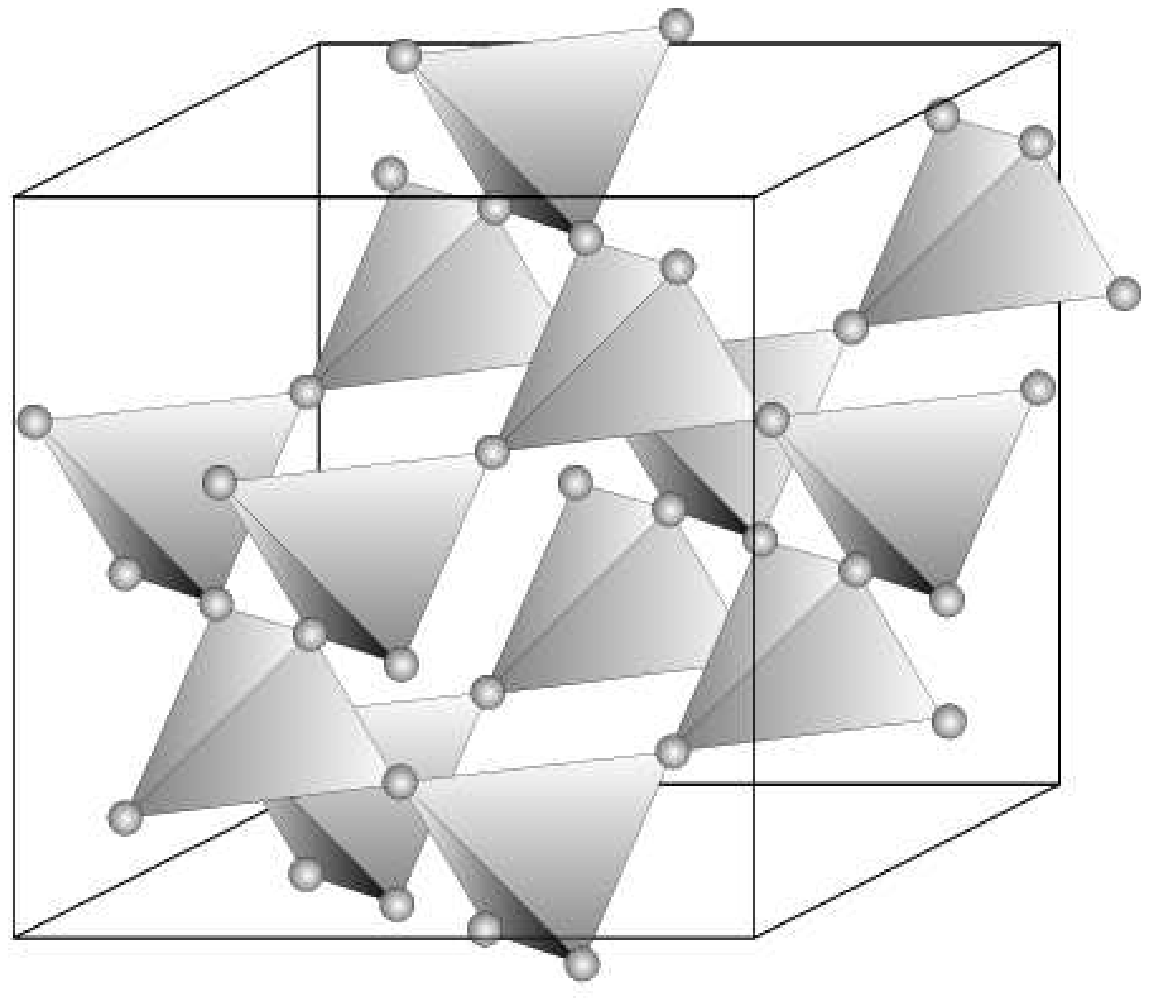}
}
\centerline{(a) \hfil (b)}
\caption{
(a) A Kagome lattice and (b) a pyrochlore lattice.
}
\label{KPlatt}
\end{figure}

\begin{figure}[tb]
\centerline{
\includegraphics[scale=0.4]{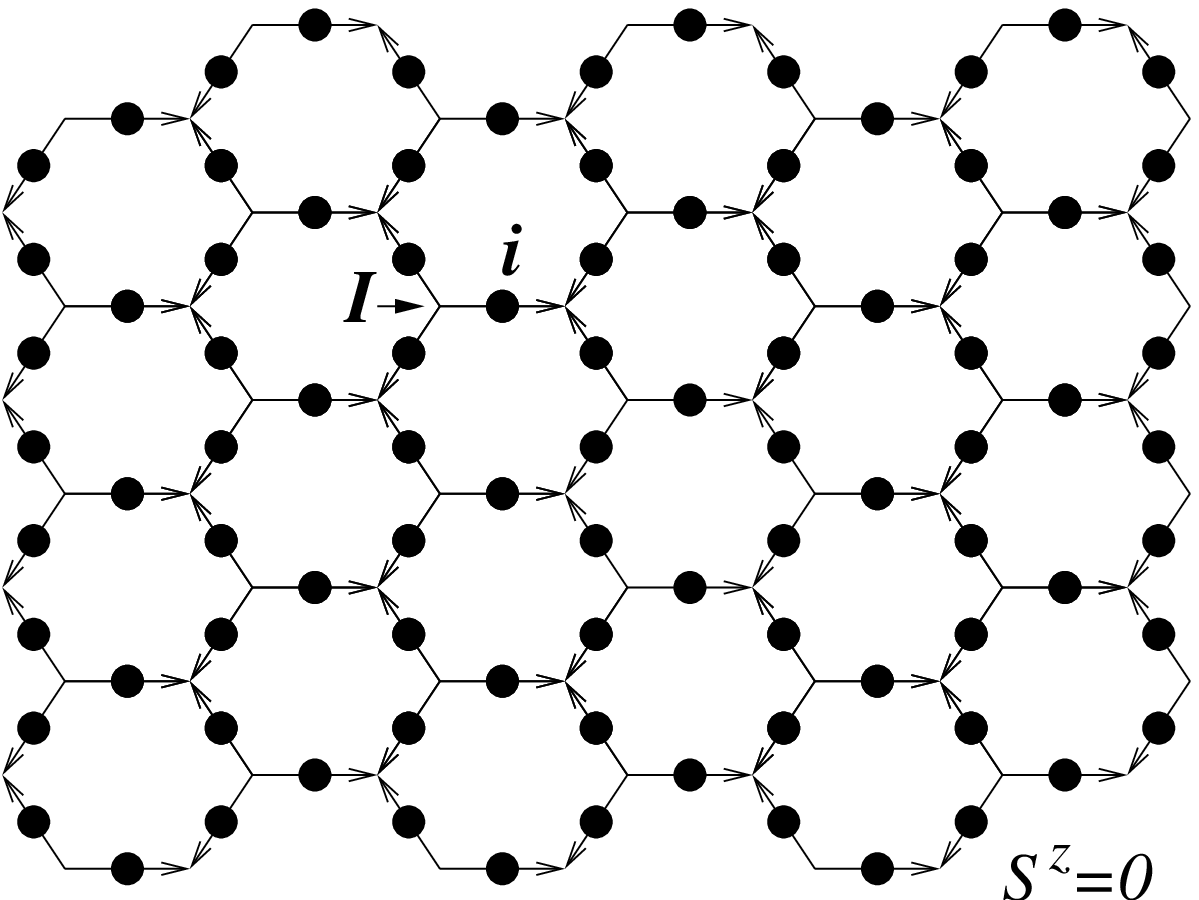}
\hfil
\includegraphics[scale=0.4]{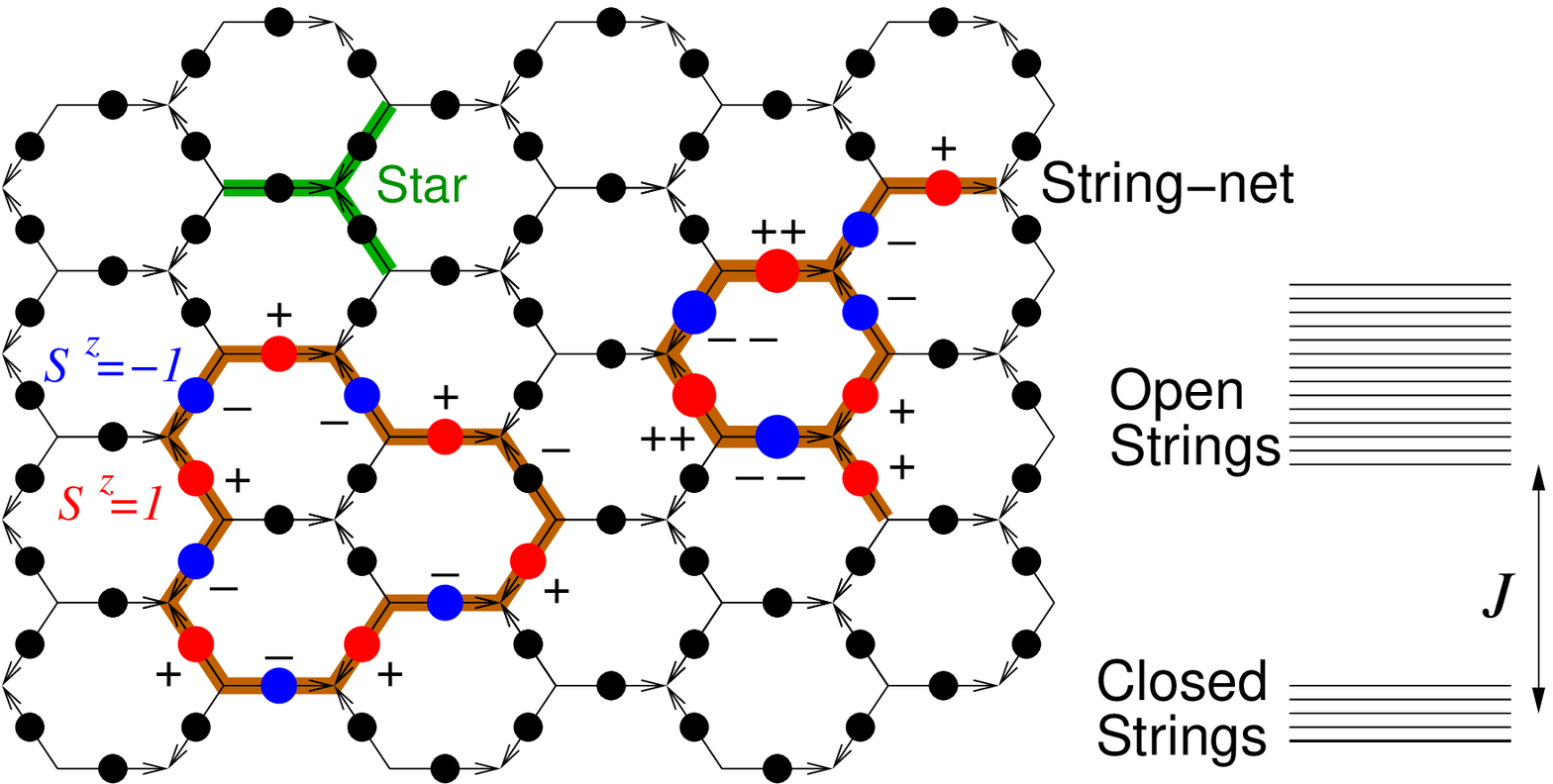}
}
\centerline{(a) \hfil\hfil (b)\hfil}
\caption{
A Kagome lattice viewed as the links of a honeycomb lattice.
(a) The no-string state, and (b) a closed string on the left
and a string-net with open ends on the right.
}
\label{KlattGS}
\end{figure}

To explain in more detail what is string-net condensation, let us consider
a XXZ spin
model on Kagome or pyrochlore lattice (see Fig. \ref{KPlatt}). \cite{Walight} The spins in the
model carry an integer angular momentum and has only on-site and
nearest-neighbor interactions:
\begin{equation*}
 H= J\sum_{\v i} (S^z_{\v i})^2
+J\sum_{\<\v i\v j\>} S^z_{\v i} S^z_{\v j}
+J_{xy}\sum_{\<\v i\v j\>} ( S^x_{\v i} S^x_{\v j} +S^y_{\v i} S^y_{\v j})
\end{equation*}
To see why the ground state of the above XXZ model is a string-net condensed
state,
let us first rewrite $H$ as
\begin{equation*}
 H= \frac J2 \sum_{\v I} Q_{\v I}^2
+J_{xy}\sum_{\<\v i\v j\>} ( S^x_{\v i} S^x_{\v j} +S^y_{\v i} S^y_{\v j})
,\ \ \ \ \ \ \
Q_{\v I} = \sum_\text{star} S^z_{\v i}
\end{equation*}
Here we have viewed the Kagome lattice as the links of the honeycomb lattice
which is labeled by $\v I$ (see Fig. \ref{KlattGS}a).  $\sum_\text{star}$ is
the sum over the three spins next to a vertex $\v I$ of the honeycomb lattice
and $\sum_{\v I}$ is the sum over all such vertices.  When $J_{xy}=0$, the
model is exactly soluble. One of the ground state is the state with all $S_{\v
i}^z=0$ which has zero energy.  We will call such a state as a no-string
state.  There are many other zero energy states. Those states can be
constructed starting from the $S_{\v i}^z=0$ state. We first draw a closed
loop in the honeycomb lattice and then alternatively increase and decrease
$S_{\v i}^z$ by $1$ along the loop.  Such a state is called a closed-string
state.  One can check that all closed-string states have zero energy and all
other states have an energy at least $J_2$. So when $J_{xy}$ is small, the low
energy states of the XXZ model are closed-string states.  When $J_{xy}\neq 0$,
the degeneracy between the closed-string states is lifted. In the small and
negative $J_{xy}$ limit, the ground state is an equal amplitude superposition
of all closed-string states.  Such a state is called a string-net condensed
state.  It represents a new state of matter. In \Ref{Walight}, it was shown
that that the collective excitations above such a string-net condensed state
behave like light and are described by Maxwell equation.

We would like to point out that the equal amplitude superposition of all
closed-strings represents the simplist string-net condensed state.  For such a
string-net condensed state, the ends of strings have a Bose statistics.
However, if we modify the $J_{xy}$ term in a certain way, the ground state of
the rotor model will correspond to more complicated string-net condensation
where the amplitudes of different string-net configuration in the ground state
may have different signs. \cite{LWqed} For such a more complicated string-net
condensed state, the ends of the strings may have a Fermi statistics.

From the simple XXZ model, we see that photons and electrons, just like
phonons, can emerge as collective motions of a proper organized particles.
Photon and electron do not have to be elementary particles, if our vacuum
itself is a string-net condensed state.  

\section{Potential applications of string-net condensation}

The materials described by Landau symmetry-breaking theory have had enormous
impact on technology. Ferromagnetic materials that break spin rotation symmetry
can be used as the media of digital information storage.  A hard drive made by
ferromagnetic materials can store so much information that books from whole
library can be put in it.  Liquid crystals that break rotation symmetry of
molecules find wide application in display.  Nowadays one can hardly find a
household without liquid crystal display somewhere in it.  Crystals that break
translation symmetry lead to well defined electronic band which in turn allow
us to make semiconducting devices.  Semiconducting devices make the high tech
revolution possible which changes the way we live.  String-net condensed states
are a new class of materials which are even richer than symmetry breaking
states.  After seeing so much impact of symmetry-breaking states, one cannot
help to imagine the possible potential applications of the richer string-net
condensed states.

One possible applications is to use string-net condensed states as media for
quantum computing. String-net condensed state is a state with complicated
quantum entanglement. As a many-body system, the quantum entanglement in
string-net condensed state is distributed among many different particles/spins.
As a result, the pattern of quantum entanglements cannot be destroyed by local
perturbations. This significantly reduces the effect of decoherence. So if we
use different quantum entanglements in string-net condensed state to encode
quantum information, the information can last much longer.\cite{DKL0252} The
quantum information encoded by the string-net entanglements can also be
manipulated by dragging the ends of strings around each others.  This process
realizes quantum computation.\cite{FKL0331}  So string-net condensed states are
natural media for both quantum memory and quantum computation.  Such
realizations of quantum memory and quantum computation are fault
tolerant.\cite{K032}

\section{Possible experimental realizations of string-net condensations}

\begin{figure}[tb]
\centerline{
\includegraphics[scale=0.5]{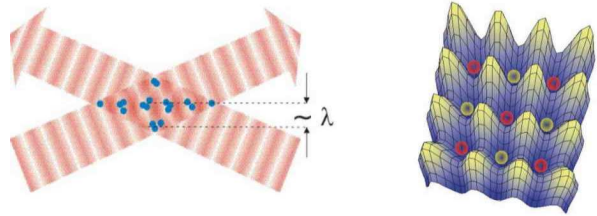}
}
\caption{
An optic lattice formed by the interference of several laser beams.
}
\label{optlatticepic}
\end{figure}

Right now, we only know how to construct theoretical models that give
rise to string-net condensations. The next step is to design realistic
materials or to find realistic materials that have string-net condensations.

We have seen that the XXZ model is a simple model with string-net condensation
which can give rise to artificial photons.  But which
experimental system can realize such a XXZ Hamiltonian?  One possibility is to
use ultra cold atoms confined in optical lattice (see Fig. \ref{optlatticepic})
to realize the XXZ model.  Usually, optical lattice is formed by the
interference of several laser beams.  But we can also use the interference of
laser beams with holographic masks to produce more complicated lattices such as
the Kagome lattice.  The optic lattice has a very useful property that the
confining potential of the optic lattice may depend on the atomic spins.  By
tuning the laster frequency strength, we can tune such a spin dependent
confining potential.  This allows us to tune the spin-spin interaction between
atoms confined in the optical lattice \cite{DDL0302} to make the XXZ
Hamiltonian.

Frustrated spin systems on Kagome lattice and pyrochlore lattice appear quite
commonly in nature.  So another possibility is to find the string-net
condensed states in those frustrated spin systems. However, the spin
Hamiltonians of those systems may not have a form of the XXZ model discussed
above.  But this does not mean that the string-net condensed states cannot
exist in those frustrated systems. String-net condensed states are stable
robust phases which exist for a finite range of parameters in the Hamiltonian.
The main difficulty is that those more general Hamiltonians are hard to solve.
Even if a Hamiltonian supports a string-net condensed ground state, we may not
know it. So it is important to develop some simple mean-field methods
for string-net condensed states. The mean-field theory will allow us to
estimate whether a Hamiltonian has string-net condensation or not. This will
allow us to determine from the measured spin interaction if a frustrated spin
system has string-net condensations or not.  We hope this line of research
will result in a list of materials which are likely to have string-net
condensations.  Experimentalists can then study those materials in detail to
check if string-net condensations really do exist in those materials or not.

The third type of promising systems is the Josephson junction
arrays.\cite{IFI0203}  The charge on a superconducting island or the flux
through a superconducting ring may play the role of spin degree freedom.
Josephson junction array is quit tunable.  Recent progress in quantum computing
has find ways to reduce the decoherence.\cite{NPT9986,VAC0286,YHC0289}  So
building a Josephson junction array to realized the quantum XXZ model may be
possible.

\section{Summary}

To summarize, string-net liquid represents a different way to understand the
deep structure of matter, where the elementary particles are not regarded as
the building blocks of everything but as an emergent phenomenon from a deeper
structure of our non-empty vacuum.  String-net liquid provides an unified
origin for almost all the elementary particles.  In other words, if we say let
there be string-net liquid, we will get almost everything.\footnote{We know
that superstring theory is a potential theory of everything. One may want to
ask what is the difference between the string-net-liquid approach and the
superstring approach?  Our understanding of the superstring theory has been
evolving.  According to an early understanding of the superstring theory, all
the elementary particles correspond to small segments of superstrings.
Different vibration modes of a small superstring result in different types of
elementary particles.  This point of view is very different from that of the
string-net liquid.  According to the string-net picture, everything comes from
simple bosonic particle.  The bosons stick together which from string-nets that
fill the whole space. The strings can be as long as the size of universe. Light
(photons) correspond to the collective motion of the large string-nets and an
electron corresponds to a \emph{single} end of string.  A modern understanding
of the superstring theory is still under development. According to Witten, one
of the most important questions in superstring theory is to understand what is
superstring. \cite{Wwhat}  So at this time, it is impossible to compare the
modern understanding of the superstring theory with the string-net theory.  In
particular it not clear if the superstring theory can be viewed as a local
bosonic model. The string-net theory is fundamentally a local bosonic model.}
This is not just a fun thing to say.  In principle, we can realize string-net
liquids in certain materials which will allow us to make artificial elementary
particles.  So we can actually create an artificial vacuum, and an artificial
world for that matter, by making a string-net liquid. This would be a fun
experiment to do!

This research was supported by NSF grant No.  DMR-0433632


\end{document}